\documentclass[conference]{IEEEtran}
\IEEEoverridecommandlockouts
\usepackage{cite}
\usepackage{amsmath,amssymb,amsfonts}
\usepackage{algorithmic}
\usepackage{graphicx}
\usepackage{graphics}
\usepackage{textcomp}

\usepackage{subcaption}
\usepackage{multirow}
\usepackage[inline]{enumitem}
\usepackage{gensymb}

\def\BibTeX{{\rm B\kern-.05em{\sc i\kern-.025em b}\kern-.08em
    T\kern-.1667em\lower.7ex\hbox{E}\kern-.125emX}}
\begin{document}

\title{TNN7: A Custom Macro Suite for Implementing Highly Optimized Designs of Neuromorphic TNNs}

% \author{\IEEEauthorblockN{Harideep Nair}
% \IEEEauthorblockA{\textit{Electrical \& Computer Engineering} \\
% \textit{Carnegie Mellon University}\\
% hpnair@andrew.cmu.edu}
% \and
% \IEEEauthorblockN{Prabhu Vellaisamy}
% \IEEEauthorblockA{\textit{Electrical \& Computer Engineering} \\
% \textit{Carnegie Mellon University}\\
% pvellais@andrew.cmu.edu}
% \and
% \IEEEauthorblockN{Santha Bhasuthkar}
% \IEEEauthorblockA{\textit{Electrical \& Computer Engineering} \\
% \textit{Carnegie Mellon University}\\
% sbhasuth@alumni.cmu.edu}
% \and
% \IEEEauthorblockN{John Paul Shen}
% \IEEEauthorblockA{\textit{Electrical \& Computer Engineering} \\
% \textit{Carnegie Mellon University}\\
% jpshen@cmu.edu}
% }

\author{\IEEEauthorblockN{Harideep Nair, Prabhu Vellaisamy, Santha Bhasuthkar, and John Paul Shen}
\IEEEauthorblockA{\textit{ECE Department, Carnegie Mellon University, \{hpnair, pvellais, sbhasuth, jpshen\}@andrew.cmu.edu}}
}

\maketitle

\begin{abstract}
Temporal Neural Networks (TNNs), inspired from the mammalian neocortex, exhibit energy-efficient online sensory processing capabilities. Recent works have proposed a microarchitecture framework for implementing TNNs and demonstrated competitive performance on vision and time-series applications. Building on these previous works, this work proposes \textit{TNN7}, a suite of nine highly optimized custom macros developed using a predictive 7nm Process Design Kit (PDK), to enhance the efficiency, modularity and flexibility of the TNN design framework. TNN prototypes for two applications are used for evaluation of TNN7. An unsupervised time-series clustering TNN delivering competitive performance can be implemented within 40 uW power and 0.05 mm\textsuperscript{2} area, while a 4-layer TNN that achieves an MNIST error rate of 1\% consumes only 18 mW and 24.63 mm\textsuperscript{2}. On average, the proposed macros reduce power, delay, area, and energy-delay product by 14\%, 16\%, 28\%, and 45\%, respectively. Furthermore, employing TNN7 significantly reduces the synthesis runtime of TNN designs (by more than 3x), allowing for highly-scaled TNN implementations to be realized.
\end{abstract}


\section{Introduction}
\label{intro}
%% TODO: 
%% Introduce macros more clearly
%% Current vs New, Baseline vs Custom
%% Make graphs instead of table
%% Can we bring back the previous layouts?
%% Make Table 1/ Fig. 1 better/more graphic?
%% Separate out synapses (including STDP), neuron body, WTA and perhaps talk about the presence of "standard" adder macro.
%% Title better?
%% Justify the results based on optimizations
%
%
%
%% Hari - The first paragraph introduces TNNs and motivates the necessity for a hardware implementation framework.
%% Hari - The second paragraph highlights the difference between direct cmos and this paper and mentions the key contributions.
Deep Neural Networks (DNNs) have 
% been at the forefront of Machine Learning for the last decade, achieving 
achieved state-of-the-art performance on diverse applications involving sensory processing tasks such as computer vision and speech recognition   \cite{lecun2015deep}. 
However, the computing demand for DNNs has been increasing exponentially and is on a highly unsustainable path in terms of computational, economic and environmental costs \cite{thompson2020computational}. In contrast, Temporal Neural Networks (TNNs) \cite{smith2018space, smith2020temporal}, a special class of Spiking Neural Networks (SNNs), strive to mimic biological neural networks with the goal of achieving both brain-like capability and brain-like energy efficiency. 

Inspired by brain's \textit{temporal} computational paradigm, TNNs are based on a rigorous \textit{space-time} algebra \cite{smith2018space} and use precise spike timings to represent and process information \cite{vanrullen2005spike}.
% that use precise spike timings to represent and process information, mimicking biological neural networks \cite{smith2018space, smith2017space, smith2020neuromorphic}.
% , which consume less power as compared to deep neural networks, owing to sparse activity and relatively simpler hardware structures.
% The key building blocks of the TNN framework are the multi-neuron columns and multi-column layers, as shown in Fig. 1 \cite{nair2020direct}. A $p\times q$ TNN column comprises \textit{q} excitatory neurons, each with \textit{p} synapses as inputs. It performs learning using Spike Timing Dependent Plasticity (STDP) learning rules and Winner Take All (WTA) inhibition. 
Unlike DNNs that utilize compute-intensive tensor processing, TNNs do not involve complex linear algebraic computation and employ simple feed-forward processing based on spikes and their timing relationships. Furthermore, TNNs are capable of online continuous learning using biologically-plausible local learning algorithms called Spike Timing Dependent Plasticity (STDP), unlike backpropagation-fueled DNNs that have a strict bifurcation between training and inference phases. 

These features make TNNs truly neuromorphic and therefore suitable for building extremely energy-efficient edge-native sensory processors for applications such as time-series clustering \cite{chaudhary2021unsupervised}.
% and capable of brain-like sensory processing and exhibits extreme brain-like energy efficiency. 
A microarchitecture framework for efficient CMOS implementation of TNNs has recently been proposed in \cite{nair2021online}. The proposed implementation methodology utilizes two notions of temporal resolution and thereby hardware clocks: 1) \textit{unit} clock serving as the finest temporal resolution to calibrate the spike timings within a single instance of input, and 2) a coarser resolution \textit{gamma} clock to separate  different input instances. The authors in \cite{nair2021online} use standard off-the-shelf 45nm CMOS to implement the key TNN microarchitectural elements,
% (illustrated in Fig. \ref{fighier})
namely, neurons, columns and STDP learning rules.
% In this proposal, standard off-the-shelf CMOS is used. This paper explores the potential of creating a custom set of macros that can be added to the cell library for more efficient implementation of TNNs.
% in CMOS.
% 
% This paper builds on and goes beyond the work in \cite{nair2021online}, 
% % which demonstrates the feasibility of implementing TNNs in CMOS using 45nm standard cell library with regular standard cells. This paper goes beyond \cite{nair2021online}, 
% exploring the potential for creating a customized cell library for further design optimization of TNNs, and can serve as the first step towards creating a scalable design framework and toolsuite for building TNN-based neuromorphic processors.
% 
% 
% \begin{figure}[t]
% \centering
% \includegraphics[width=\columnwidth]{diagrams/fig1_hierarchy.png} 
% \caption{Temporal Neural Network Hierarchy: Synapses, Neurons, Columns and Layers}
% \label{fighier}
% \end{figure}

This paper builds on and goes beyond the work in \cite{nair2021online}, exploring the potential for creating a customized cell library that utilizes inherent TNN principles to improve the power, performance and area (PPA) of TNN designs. Furthermore, this work  serves as the first step towards creating a scalable design framework and toolsuite for building TNN-based neuromorphic processors.
We make four key contributions: 1) the TNN design process, including gate-level implementations, is replicated in 7nm predictive CMOS using the ASAP7 Process Design Kit (PDK) \cite{clark2016asap7}, and post-synthesis Power-Performance-Area (PPA) results are reported; 2) a set of nine new highly-optimized custom macro extensions to ASAP7, called \textit{TNN7}, that can be used for implementing highly energy-efficient parameterizable TNNs is proposed; 3) significantly improved scaling of PPA as well as synthesis runtime for larger design sizes, achieved by TNN7, is demonstrated; and 4) the hardware complexities of TNN prototypes in \cite{smith2020temporal} 
for image classification, and \cite{chaudhary2021unsupervised} for unsupervised time-series clustering, are evaluated and shown to achieve significant improvements using the custom macro extensions, demonstrating the potential of TNNs for energy-efficient sensory processing with online learning.
% 4) an added efficacy for realizing TNN7 custom library is demonstrated by illustrating significant improvements in synthesis runtime by comparing the runtimes of ASAP7-based TNN columns and TNN7-incorporated columns used for the UCR time-series clustering. 
%Our results can serve as baseline benchmarks for future works. 
% A set of 11 macros are proposed - \textit{syn\_weight\_update}, \textit{syn\_output, pac\_adder}, \textit{less\_equal}, \textit{pulse2edge}, \textit{stdp\_case\_ gen}, \textit{stabilize\_func}, \textit{incdec}, \tetxit{mux2to1gdi}, \textit{edge2pulse}, and \textit{spike\_gen}. 
% We illustrate their utility by implementing TNNs comprising entirely with the novel cell macros and note the optimization gains across key hardware metrics. 
% 
% \section{Temporal Neural Networks}
% \label{tempnn}
% % Hari - TODO: This section needs a complete rewriting.
% % Subsections - 1) Organization and 2) uarch framework
% % Prabhu - changed the section, have a look? Hari - Thanks! Will go through.
% 
% %% Hari (02/20) - I am yet to work on this Section. Will finish it up by tonight
% 
% \begin{figure}[t]
% \centering
% \includegraphics[width=\columnwidth,height=5.7cm]{diagrams/fig1_hierarchy.png} 
% \caption{Hierarchical TNN Organization 
% % (from \cite{nair2020direct})
% }
% \label{fig:tnn_org}
% \end{figure}
% \vspace{-10pt}
% 
\begin{figure}[t]
\centering
\includegraphics[width=\columnwidth]{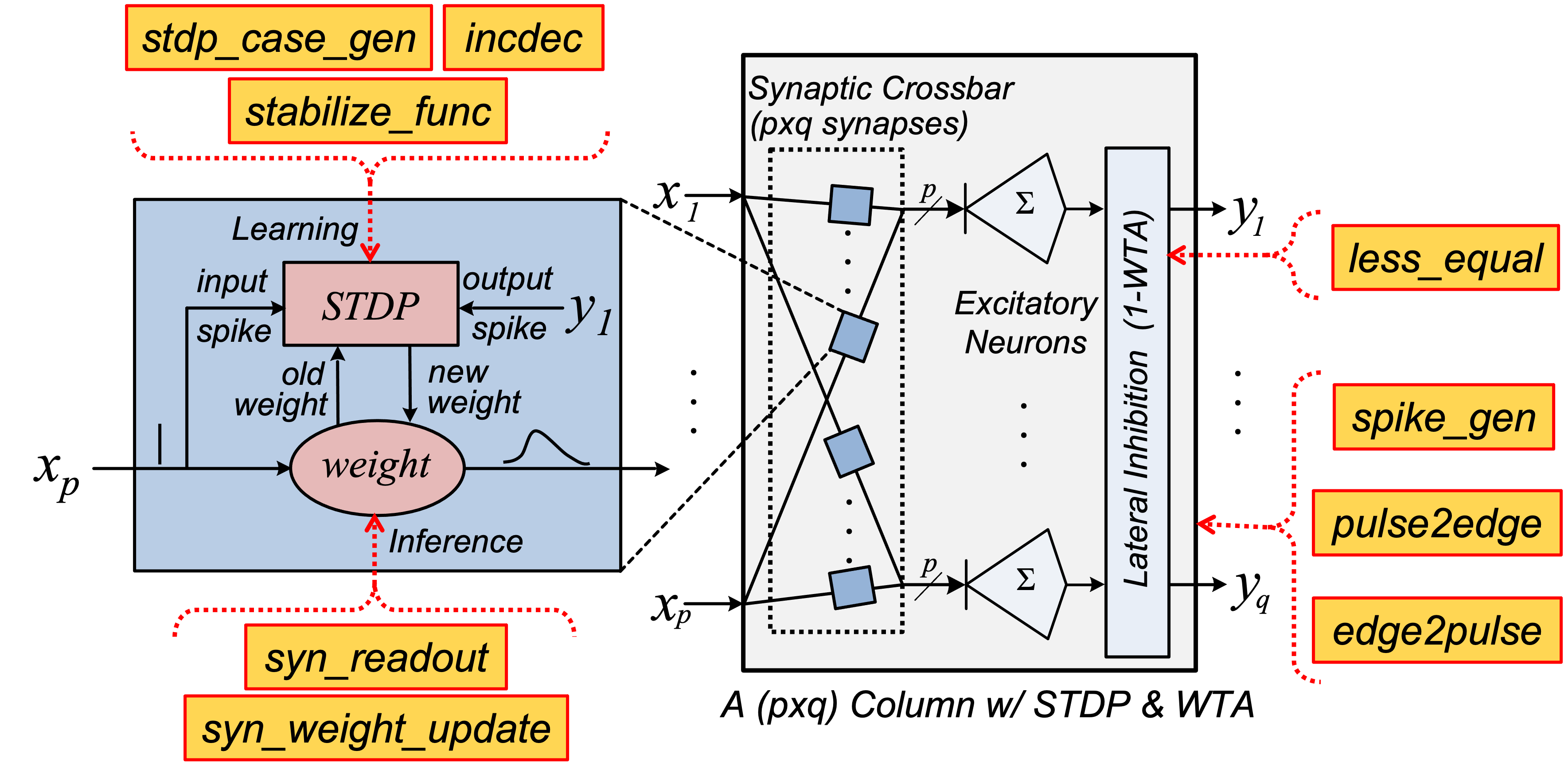} 
\caption{Functional components of a \textit{p}x\textit{q} TNN Column and associated custom macros (highlighted in yellow)}
\label{macro_bd}
\end{figure}
\begin{table}[t]
\centering
\caption{Proposed Custom Macros}
% \scalebox{0.97}{
\resizebox{\columnwidth}{1.9cm}{
 \begin{tabular}{|c|c|c|l|} 
 \hline
 TNN & Proposed & Function & Figure \\
 Units & Macros & Description & Label \\
 \hline
 \hline
 Synaptic & \emph{syn\_readout} & Perform RNL readout & Fig. \ref{fig:fsm_output_f} \\
 \cline{2-4}
 Response & \emph{syn\_weight\_update} & Perform weight update & Fig. \ref{fig:syn_weight_update_f} \\
 \hline
 WTA & \emph{less\_equal} & Perform temporal inhibit & Fig. \ref{fig:inhibit_f} \\
 \hline
 \multirow{3}{*} {STDP} & \emph{stdp\_case\_gen} & Control STDP cases & Fig. \ref{fig:stdp_f}\\
 \cline{2-4}
 & \emph{incdec} & Control update direction & Fig. \ref{fig:incdec_f} \\
 \cline{2-4}
  & \emph{stabilize\_func} & Stabilize weights bimodally & Fig. \ref{fig:stabilize_func_f} \\
 \hline
 \multirow{3}{*} {Utility} & \emph{spike\_gen} & Perform spike encoding & Fig. \ref{fig:spike_gen} \\
 \cline{2-4}
 & \emph{pulse2edge} & Convert from pulse to edge & Fig. \ref{fig:pulseP_f} \\
 \cline{2-4}
  & \emph{edge2pulse} & Convert from edge to pulse & Fig. \ref{fig:edge_f} \\
 \hline
 \end{tabular}
 }
  \label{tab:custom_macros}
\end{table}

% \vspace{-12pt}
The proposed nine macros have been designed to target and optimize the primary TNN building blocks, or TNN columns.
% and provide the capability to model arbitrary TNN designs.
% (as in Fig. \ref{fighier}) with arbitrary number of synapses, neurons, columns and layers. 
% 
% They are described in detail in the next section. 
% Fig. \ref{fig:tnn_org} shows the overall hierarchy of a multi-layer TNN with columns as the key building blocks, and 
Fig. \ref{macro_bd} illustrates the custom macros in a typical \textit{p}x\textit{q} column (key TNN building block) with \textit{p} synapses per neuron and \textit{q} such neurons, followed by winner-take-all (1-WTA) lateral inhibition.
As majority of the TNN computation occurs in the synaptic crossbar, five macros are dedicated for synapses (two for synaptic inference or response function generation and three for STDP local learning). These synapses then feed into corresponding neuron bodies which perform response function summation through adder trees. Another macro enables the key comparison operation in WTA and the remaining three macros serve more generic utility purposes (e.g. spike encoding).

These macros (elaborated in Section \ref{macrosec}) are summarized in Table \ref{tab:custom_macros} along with their functional descriptions and schematic figure labels.
It should be noted, these custom macros can be generalized beyond use in the 
% particular learning rules or neuron 
microarchitecture model in \cite{nair2021online}, and
can serve as the foundation for building generic temporal functions based on \textit{space-time} algebra \cite{smith2018space}.
% can serve as CMOS primitives for implementing generic temporal functions based on \textit{space-time} algebra \cite{smith2018space}.
To the best of our knowledge, this is the first work that proposes custom macros for highly efficient scalable CMOS implementation of TNNs.
\section{Design Framework \& Methodology}
% This line should indicate this is a direct sequel to ISVLSI.
% Enhancing the applicability and efficiency of the TNN microarchitecture in \cite{nair2021online} is the focus of this work. This involves developing an optimized custom macro library suite from the fundamental building blocks that constitute TNNs. A scalable TNN design framework can be realized using these custom macros, with the benefits of PPA optimizations achieved from using these custom macros.
%This section describes the circuit-level designs of the proposed custom macros and their function in detail. Also, the CAD flow used for their design are discussed.
% tools and the PDK used for our design flow are discussed.
% and contribution to the design of the TNN framework.
The proposed TNN7 custom cells are developed as hard macros using an open-source 7nm predictive Process Design Kit (PDK), called ASAP7 \cite{clark2016asap7}. This section describes the ASAP7 library and the CAD design flow used in this work.
\subsection{Framework}
\label{framework}
ASAP7 \cite{clark2016asap7} is an academically certified, foundry agnostic, predictive PDK based on 7nm finFET technology. This involves a standard cell library and a collection of rule-sets for physical verification - design rule checks, layout vs. schematic, and parasitic extraction. The electrical activity of the transistor models is scaled from the BSIM-CMG SPICE models \cite{duarte2015bsim}, which captures advanced trends in the finFET industry. ASAP7 offers transistor device models at four threshold voltages (SLVT, LVT, RVT and SRAM), and three process corners, typical-typical (TT), slow-slow (SS) and fast-fast (FF). 

% DO WE NEED TO EXPLAIN WHY THESE CHOICES?????
In this work, following selections are used for the design of custom macros:
1) RVT device models with nominal operating conditions at TT corner (0.7V supply voltage and 25\degree C operating temperature),
2) composite current source (CCS) modeling for timing files, and
3) Cadence/Mentor Graphics toolchain for logic synthesis, schematic, layout and characterization. 
% Using this setup, hard macros are constructed with aggressive PPA optimizations.
% 3) Cadence toolchain - \textit{Genus} for register-transfer level (RTL) logic synthesis, \textit{Virtuoso} for schematics and layouts, \textit{Liberate} for characterization of the custom cells, and \textit{Abstract} for the development of liberty exchange format (LEF) files.
% for implementing fundamental functional modules needed for designing TNNs.
\subsection{Methodology}
\label{methodology}
% Developing an optimized custom macro library suite from fundamental functional blocks that constitute a Temporal Neural Network is the chief contribution of this work. To establish the improvement in the PPA metric numbers achieved by incorporating the custom macros, the following methodology is followed to report the results of this work:
% This section describes the use of Cadence toolchain and Mentor Calibre decks in developing the custom macros and presents the methodology followed for reporting the optimization gains.
% This section describes the methodology used for developing the custom macros and reporting the optimization gains presented in Section \ref{benchmark}.

%Using the framework described above, the following methodology is adopted to develop custom hard macros and report their PPA results and improvements.

In developing the custom macros, Cadence tool suite is used as follows: 1) \textit{Genus} for register-transfer level (RTL) logic synthesis, 2) \textit{Virtuoso} for schematics and layouts, 3) \textit{Liberate} for characterization of the macros and generating Liberty (.lib) timing files, and 4) \textit{Abstract} for generating Liberty Exchange Format (.lef) files of the macros. Layout verification, including Layout Versus Schematic (LVS) and Design Rule Check (DRC), is performed using Mentor Calibre and the resulting LVS \& DRC-clean Graphic Data Stream (GDS) files are imported to \textit{Abstract}. Moreover, Calibre Parasitic Extraction (PEX) tool reads the layout and generates the extracted netlist which is then used for Spectre simulations in \textit{Liberate}.

In order to report the optimization gains presented in Section \ref{benchmark}, following steps are adopted: 1) \textit{Genus} is used to synthesize the original functional modules from \cite{nair2021online} with the ASAP7 standard cell library and establish the baseline values; 2) TNN7 macro equivalent of the original modules are designed by either (i) structurally optimizing at the microarchitectural level, or (ii) creating mixed-signal circuits from scratch in \textit{Virtuoso}; 
% Consequently, \textit{Liberate} and \textit{Abstract} tools are employed to generate the custom .lib and .lef files for each macro. 
3) \textit{Genus} is used to resynthesize the modules by replacing the ASAP7 standard cells with the TNN7 .lib and .lef files (obtained from \textit{Liberate} and \textit{Abstract}), to obtain post-synthesis area, power and delay. These values are then compared against the ASAP7-based post-synthesis values to compute the corresponding improvements.
% , and improvements for individual modules are computed.
% Finally, a comprehensive PPA metric comparison analysis between the standard cell-based column design \cite{nair2020direct} and the custom macro incorporated column design is provided in Section IV. This work can be envisioned as a neuromorphic library extension to the ASAP7 standard cell library specifically for constructing TNNs.

% The TNN framework can be realized entirely from the comprehensive list of custom macros, with the additional benefits of aggressive Power, Performance, Area (PPA) optimizations achieved from the custom cell suite.
% \subsection{Optimization Strategies}
% Since TNNs are being envisioned for online, always-on, and edge-native sensory processing units operating at real-time kHz frequencies, PPA optimizations are prioritized for first reducing power consumption followed by area and delay. We use Gate Diffusion Input (GDI)-based designs for maximal area and power optimizations, since they provide fast, low-power circuits with reduced number of transistors \cite{morgenshtein2001gate}. However, the tradeoff is degraded output levels which is corrected by using level restorers at the outputs. Diffusion sharing between gates in the design is used to further reduce the area.
% \vspace{-8pt}
\section{TNN7 Custom Macro Cells}
\label{macrosec}
This section describes the circuit-level design of the proposed nine macros and their functionalities in detail. 
% Each macro is implemented separately and integrated into the TNN7 library using the corresponding .lib, .lef and .gds files.
The macros are segregated into \textit{TNN functionality cells}, that perform exclusive TNN functions, and \textit{utility cells}, that perform generic functions like spike encoding. 
% The macros are segregated into those that perform exclusive TNN functions called \textit{TNN functionality cells}, and those, named \textit{utility cells}, performing more generic functions like spike encoding. 
% For each proposed custom macro, its PPA values are reported in Table \ref{tab:allmacros}.
% We describe the TNN7 library as consisting of .lib, .lef and .gds files of all the individual macros described in this section. Further, the library is segregated as into column-specific cells or utility cells. We elaborate on the functionality of each of the individual macros in detail as follows- 

\subsection{TNN Functionality Cells}
% \vspace{-3pt}
%Developing an optimized custom macro cell suite from fundamental functional blocks that constitute a Temporal Neural Network has been the chief contribution of this work. 
% 
%This work's chief contribution is distinguishing and developing an optimized custom macro cell suite out of the fundamental functional blocks that constitute a TNN framework. 
%Firstly, the TNN framework is replicated in the ASAP7 process design flow, providing the baseline 7nm PPA values of the individual sub-modules post-synthesis. Subsequently, after developing the equivalent standard cells for the same using the ASAP7 PDK and employing the Cadence toolchain, we compare and obtain the optimization gained. 
This subsection describes the six macros implementing synaptic response, WTA and STDP.
% native to the column layers capable of performing unsupervised clustering in the TNN framework. 
% The original ASAP7 framework modules are referred to as \textit{standard cells}, and proposed novel modules as \textit{custom macros}.
% along with those of the corresponding standard cells.
The following notations are used for the two hardware clocks introduced in Section \ref{intro} - \textit{aclk} for the unit clock and \textit{gclk} for the gamma clock. 

\begin{figure}[!ht]
\centering
\includegraphics[width=0.9\columnwidth]{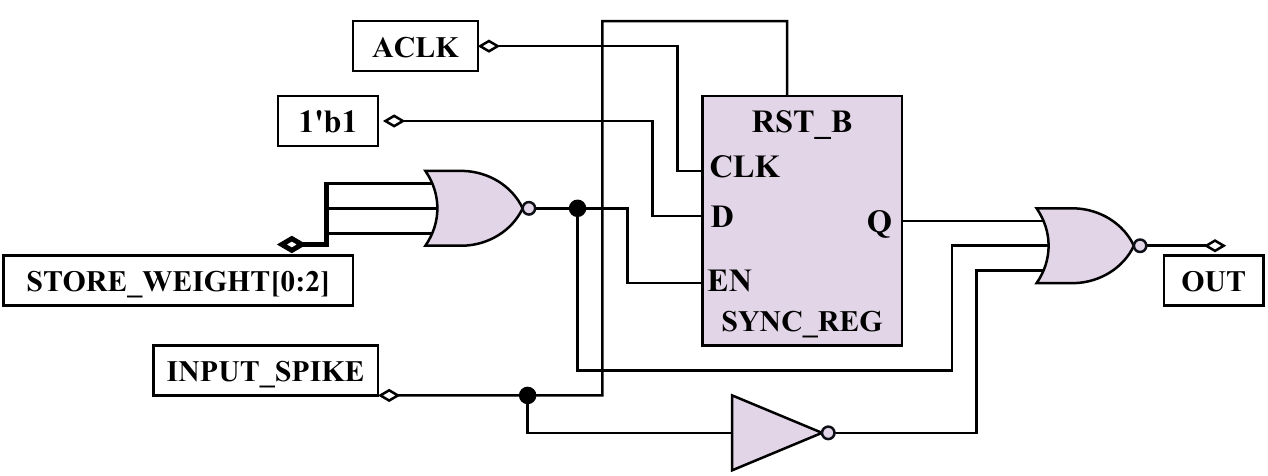} 
\caption{\emph{syn\_readout} macro}
\label{fig:fsm_output_f}
\end{figure}
\begin{figure}[!ht]
\centering
\includegraphics[width=\columnwidth]{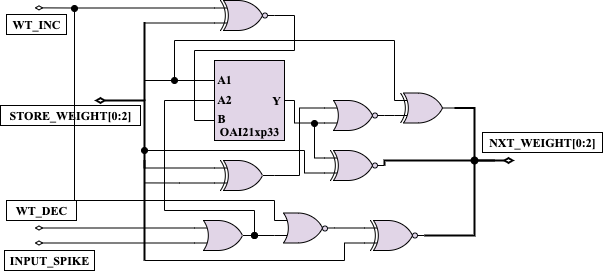} 
\caption{\emph{syn\_weight\_update} macro}
\label{fig:syn_weight_update_f}
\end{figure}

\subsubsection{syn\_readout and syn\_weight\_update}
% Prabhu - modified; need to explain how we remapped the logic to get optimization. Santha can add this part? 
%The Syanptic weight implementation is  hardware-intensive logic block to try and characterize into a custom macro. The finite state machine (FSM)-based logic implements the storage of the synaptic weight and readouts into a unary corresponding to the RNL response function. The output remains asserted until the weight is counted down to zero, after which it is reset. \\

As noted in \cite{nair2021online}, synapses constitute majority of the hardware complexity in TNNs.
% is attributed to synapses. 
Hence, in this work, main synaptic functions are identified, optimized and modularized into custom macros. 
% consitute most of the hardware complexity in TNNs, as illustrated in \cite{nair2020direct}. 
% Synapses are meticulously built FSM counters which can not only generate response functions but also store and update weights. 
% Each synapse can be broken down into two distinct components: 1) flip-flops that store the weight bits, and 2) the associated control logic that performs weight readout and update. The complex control logic 
% Synapses are meticulously designed counters with two key functionalities, namely, response function generation (or readout) and weight update.
The two key synaptic functions of response function generation and weight update are implemented as \emph{syn\_readout} (Fig. \ref{fig:fsm_output_f}) and \emph{syn\_weight\_update} (Fig. \ref{fig:syn_weight_update_f}) macros respectively. When an input spike pulse arrives, the synaptic weight undergoes a unit decrement every cycle, until it wraps around to the original value. During this process, the \emph{syn\_readout} macro takes in the weight value every cycle and asserts the output until the weight reaches zero, and then deasserts it. This parallels the unary-coded ramp-no-leak (RNL) response function in \cite{nair2021online}.
% output is nothing but the unary-coded ramp-no-leak (RNL) \cite{nair2021online} response function.
% , which eventually gets accumulated into the membrane potential in the neuron body. 
The \emph{syn\_weight\_update} macro controls the weight decrementing process during readout, and updates the synaptic weight during ``learning'', via the STDP-based control signals (\textit{WT\_INC} and \textit{WT\_DEC}). Only one of the control signals is active at a time and performs either unit increment or decrement. Note that the \emph{syn\_weight\_update} macro merely updates the weight based on external control signals; the control signals are generated by input spike (inference) and the three STDP macros (learning).

This modular approach to designing synapses provides flexibility to implementing TNN frameworks. For example, the response function can be changed by modifying \emph{syn\_readout} while keeping the other macros intact. This flexibility adds to the diversification of TNN models for diverse applications.

\begin{figure}[!ht]
\centering
\includegraphics[width=0.6\columnwidth]{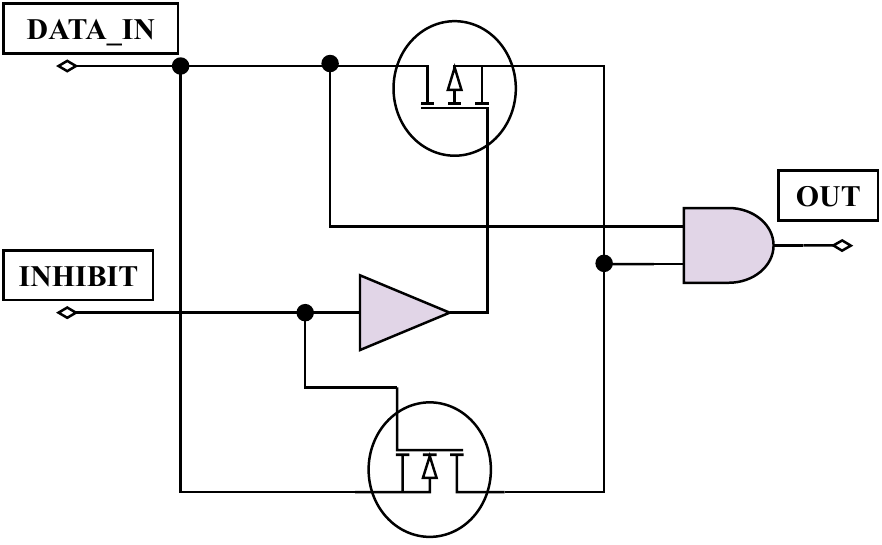} 
\caption{\emph{less\_equal} macro}
\label{fig:inhibit_f}
\end{figure}

\subsubsection{less\_equal}

The \emph{less\_equal} macro (Fig. \ref{fig:inhibit_f}) models the temporal inhibitor and functions as the basic unit for WTA inhibition. More generally, it implements the temporal operation of ``less\_equal'' from space-time algebra \cite{smith2018space} and hence is widely used in the TNN design framework. The input data (DATA\_IN) value is propagated to the output if and only if it arrives earlier or at the same time as INHIBIT; else, it is suppressed.
%(JPS) The inhibition functionality is achieved by the use of just a single transistor \cite{tzimpragos2019boosted}. However, to mitigate the high leakage current observed during the cell's characterization, it is coupled with another transistor to make NMOS-PMOS pair.
% The \emph{less\_equal} macro models the key function of the Winner-Take-All (WTA) inhibition and acts as a temporal inhibitor that enables the input data (DATA\_IN) to be transferred to the output only if it arrives earlier or at the same time as the inhibit input (INHIBIT). More generally, it implements the temporal operation of ``less\_equal'' from space-time algebra \cite{smith2018space} and hence is widely used in the TNN design framework.
% It is one of the macros that delivers the most PPA benefits as compared to standard cells.
% 
% transpiring in the biological brain, and hence is ubiquitous across the general TNN framework. 
% It functions as a temporal inhibitor that enables the input data to be transferred if and only if it arrives earlier or at the same time as the inhibit input. 
This module's functionality can be achieved by using a single transistor \cite{tzimpragos2019boosted}. However, to mitigate the high leakage current observed during the cell's characterization, a pair of NMOS and PMOS transistors is employed. 

\begin{figure}[!ht]
\centering
\includegraphics[width=0.84\columnwidth]{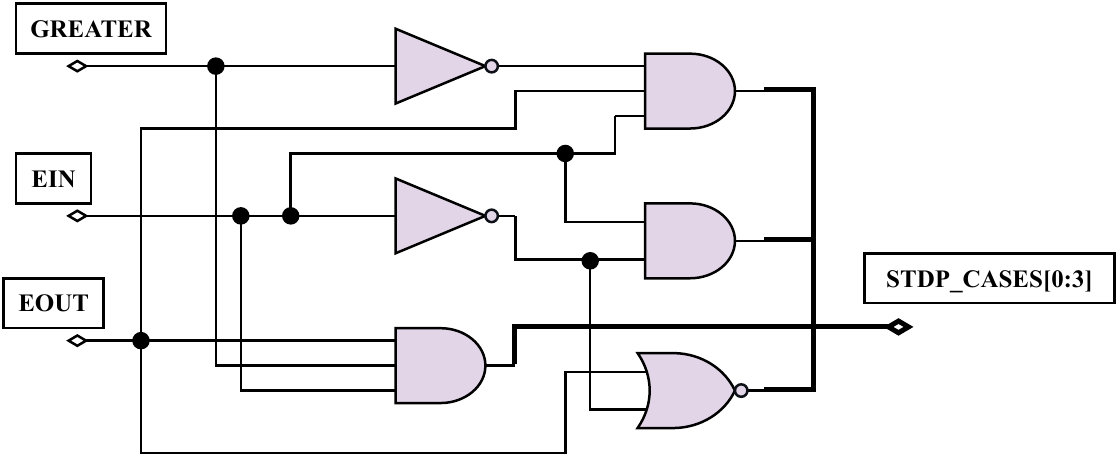} 
\caption{\emph{stdp\_case\_gen} macro}
\label{fig:stdp_f}
\end{figure}

\subsubsection{stdp\_case\_gen}
The \emph{stdp\_case\_gen} macro generates the essential control signal outputs, corresponding to the four STDP cases from Table I in \cite{nair2021online}. As shown in Fig. \ref{fig:stdp_f}, it takes in the negated output of \emph{less\_equal} (GREATER) and input/output spikes represented as edge transitions (EIN/EOUT), and generates a one-hot encoded output for the STDP cases.
% of \emph{capture}, \emph{minus}, \emph{search} and \emph{backoff}. 
% These cases, in turn, produce the relevant Bernoulli Random Variables (BRVs). 
% Figure \ref{fig:stdp_f} depicts the combinational logic that achieves the STDP learning cases' aforementioned functionality.
When both input and output spikes are absent, the output is zero, resulting in no weight update during STDP. 
% Cell area and leakage power reduce by 47\% and 68\% respectively, with an improvement of 3 ps for the delay (Table \ref{tab:allmacros}).

% \begin{table}[!ht]
% \centering
% \caption{Standard vs. custom 7nm PPA for \textit{stdp\_case\_gen}}
% \scalebox{1.1}{
%  \begin{tabular}{|c|c|c|c|} 
%  \hline
%  Design & Cell Area & Leakage Power & Delay \\
%   & ($\mu m^2$) & ($nW$) & ($ps$)\\
%  \hline
%   Standard & 17.96 & 1.13 & 73 \\ 
%   \hline
%   Custom & 9.56 & 0.36 & 70 \\
%  \hline
%  \end{tabular}}
% %  \vspace{2 mm}
%   \label{tab:stdp_case_gen}
% \end{table}
%

\begin{figure}[!ht]
\centering
\includegraphics[width=0.86\columnwidth]{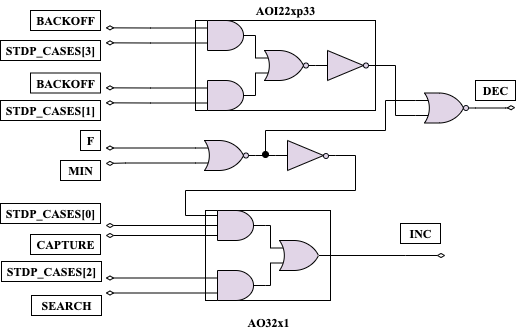} 
\caption{\emph{incdec} macro}
\label{fig:incdec_f}
\end{figure}

\subsubsection{incdec}
The \textit{incdec} (Fig. \ref{fig:incdec_f}) macro takes in the STDP cases and Bernoulli random variables (BRVs) as inputs (as in \cite{nair2021online}), and generates control signals for driving the local synaptic weight update process. It consists of AND-OR-INVERT (AOI) cells that activates INC for STDP cases 0 and 2, and activates DEC for cases 1 and 3, if the BRV is one. It is important to note that the modularity in STDP logic (due to \emph{stdp\_case\_gen} and \emph{incdec}) allows for easy modification of STDP rules.

\begin{figure}[!ht]
\centering
\includegraphics[width=0.8\columnwidth]{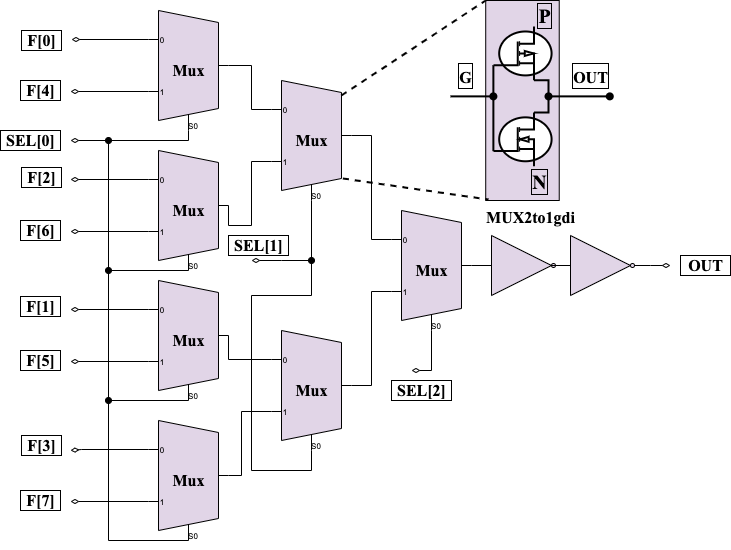} 
\caption{\emph{stabilize\_func} macro}
\label{fig:stabilize_func_f}
\end{figure}

\subsubsection{stabilize\_func}
%As the name suggests, this macro (Fig.\ref{fig:stabilize_func_f}) is responsible for selecting the appropriate BRVs using the synaptic weight as per the stabilization function to stabilize weight convergence. Our contribution is remapping the structure in \cite{nair2020direct} to an 8:1 multiplexer-based architecture, where depending on a particular synaptic weight, it chooses the corresponding value with the expected probability. Further optimization is realized by replacing the standard cell-based multiplexer design with a custom Gate Diffusion Input (GDI)-based 2:1 multiplexer \cite{morgenshtein2001gate}, and using for the \textit{stabilize\_func} synthesis. GDI cells replicate conventional Boolean arithmetic with only two transistors, enabling them to serve as fast, low-powered circuits. The GDI multiplexer is developed using the ASAP7 transistor models, performing custom schematic and layout consequently. However, the tradeoff here is degraded output levels, corrected by applying level restorers at the outputs. \\
%To summarize, the standard cell-based design is remapped to seven 2:1 GDI-based multiplexers interconnected to make the combinational block behave functionally equivalent to that of a single 8:1 combinational multiplexing logic Fig.

This macro (Fig. \ref{fig:stabilize_func_f}) is responsible for selecting the appropriate BRVs as per the stabilization function in \cite{nair2021online}, and plays a key role in establishing weight convergence.
It is architected as an 8:1 multiplexer module with a hierarchy of Gate Diffusion Input (GDI) cells \cite{morgenshtein2001gate}, each acting as a 2:1 multiplexer. The 2:1 GDI multiplexers 
utilize just two transistors, however suffer from degraded output levels. This is corrected by applying level restorers at the output, making the final design both robust and highly efficient.

\subsection{Utility Cells}
% The remaining three macros are detailed in the following subsections. These utility custom cells are generalized for the broader TNN framework, and not specific to the column layers.
The remaining three macros are utility cells generalized to perform broader functions within the TNN framework such as spike encoding, synchronization, etc. They are detailed below.

\subsubsection{spike\_gen}
%this macro (Fig.\ref{fig:fsm_simple_f}) is used to generate 8-cycle wide pulses for spikes required by \emph{spike\_gen}, and \emph{edge2pulse} macros to generate reset pulses (grst) from gclk for performing essential computational reset between consecutive computational cycles

\begin{figure}[!ht]
\centering
\includegraphics[width=\columnwidth]{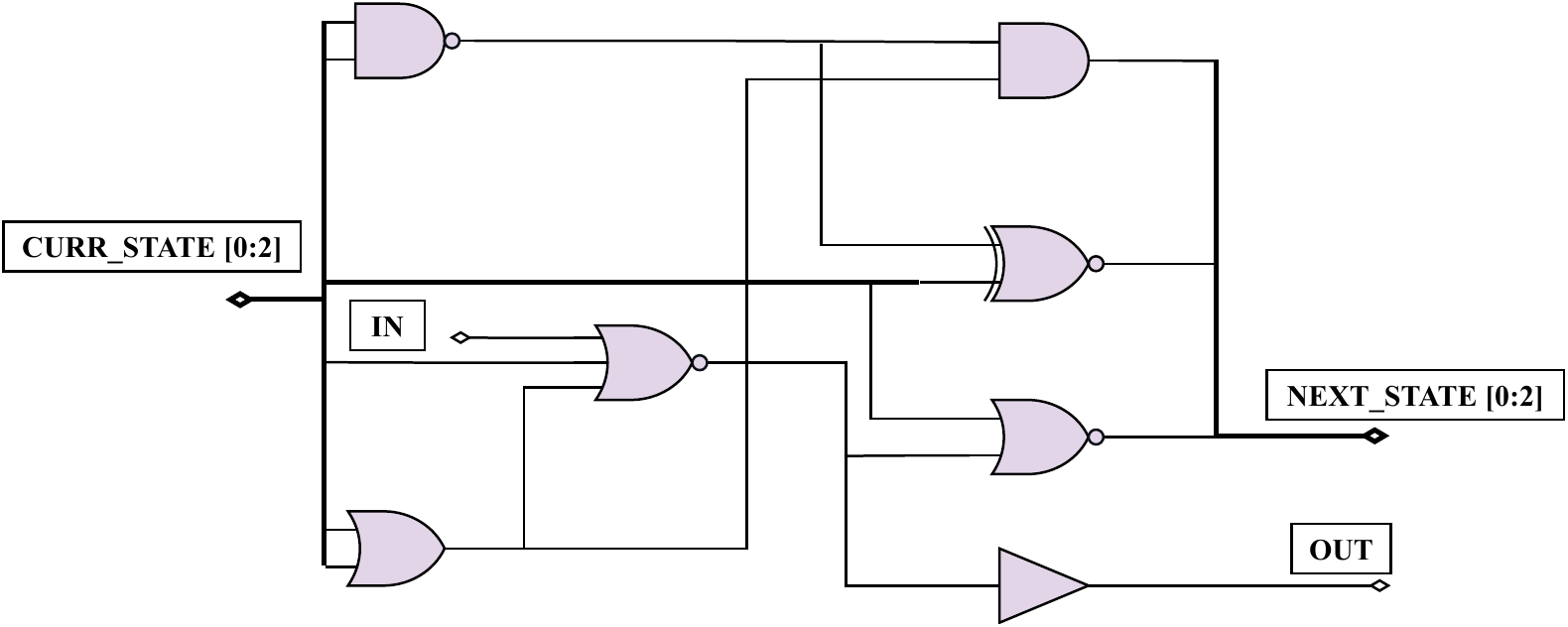} 
\caption{\emph{spike\_gen} macro}
\label{fig:spike_gen}
\end{figure}

The \emph{spike\_gen} macro (Fig. \ref{fig:spike_gen}) plays a key role in spike encoding. It implements the combinational logic associated with a 3-bit counter used to convert input pulses of any width to an 8 cycle-wide output pulse (for 3-bit synaptic weights). As demonstrated in \cite{nair2021online}, this spike encoding is central to the ramp-no-leak (RNL) functionality of the compact synapse design used in the TNN framework, and can be easily extended to generalize for arbitrary pulse widths.
%(JPS) Table \ref{tab:allmacros} reports notable improvements in all three PPA metrics, with 20\%, 30\%, and 36\% reduction in area, leakage power and delay respectively compared to the standard design.
% As illustrated in Figure \ref{fig:fsm_simple_f}, the \emph{spike\ gen} macro encompasses a combinational logic that implements the FSM logic to produce an eight clock cycle-wide output for a given arbitrary clock cycle input. It generates an input spike that is $W_{max}+1$ \textit{aclk} cycles-wide, and guides the \emph{syn\_ weight\_update} macro functionality during the readout phase. The macro is an essential constituent of the RNL response function of the neuron.

\subsubsection{pulse2edge and edge2pulse}
\begin{figure}[!ht]
\centering
\includegraphics[width=0.65\columnwidth]{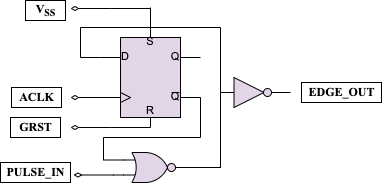} 
\caption{\emph{pulse2edge} macro}
\label{fig:pulseP_f}
\end{figure}

\begin{figure}[!ht]
\centering
\includegraphics[width=0.81\columnwidth]{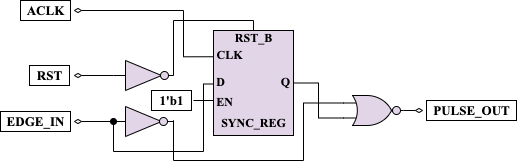} 
\caption{\emph{edge2pulse} macro}
\label{fig:edge_f}
\end{figure}

%\emph{Pulse2edge} \emph{pulse2edge} macros are developed to convert pulses to edges (in representing spikes), i.e., to stay asserted until a gamma reset. Two variants, namely, power-optimized (contains asynchronous
%active high reset register) as well as area optimized (contains synchronous active low reset register) are constructed. \emph{edge2pulse} performs the converse, producing input spikes upon receiving pulse signals.
TNN implementations utilize edge-encoded signals (i.e., encoded as edge transitions from 0$\rightarrow$1) for performing various temporal operations. The \emph{pulse2edge} macro (Fig. \ref{fig:pulseP_f}) transforms an incoming pulse signal into an edge signal (lasting until the end of current \textit{gclk} cycle), and is used extensively across the TNN framework. On the contrary, \textit{edge2pulse} macro (Fig. \ref{fig:edge_f}) outputs a pulse lasting one \textit{aclk} cycle as soon as an edge signal arrives at its input. 
It is typically used to produce internal reset pulses from \textit{gclk} to synchronize the sequential blocks in the datapath.

All nine macros have been carefully designed to use minimal number of gates and transistors to achieve their corresponding functionalities.
In order to further reduce cell area, we perform diffusion layer overlapping during manual layout. Table \ref{tab:allmacros} reports their respective PPA metrics.
% for each of the nine macros above.
% Table \ref{tab:allmacros} reports the PPA metrics for the designs of the custom macros in 7nm CMOS. 
% A common optimization technique applied across all proposed macros is diffusion layer overlapping during manual layout, which reduces cell area. 
In order to demonstrate their benefits, these macros are used to build various TNN prototypes as discussed in the next section.
% As per Table \ref{tab:allmacros}, \emph{pulse2edge} macro improves delay and area by 7\% and 9\% respectively, with a significant tradeoff in leakage power which increases by 29\%. In contrast, \textit{edge2pulse} improves leakage power, delay and area by 53\%, 26\% and 35\% respectively. Note that all proposed macros at least improve two of the three PPA metrics, wherever tradeoff becomes necessary.
% , excluding the synaptic weights.

\begin{table}[!ht]
\centering
\caption{7nm PPA for proposed custom macros}
% \scalebox{0.95}{
\resizebox{\columnwidth}{1.8cm}{
 \begin{tabular}{|c|c|c|c|} 
 \hline
 Custom Macro & Leakage Power & Delay & Cell Area \\
 Name & ($nW$) & ($ps$) & ($\mu m^2$) \\
 \hline
 \hline
 \textit{syn\_readout} & 0.43 & 32 & 0.50\\
 \hline
 \textit{syn\_weight\_update} & 1.22 & 190 & 1.24\\
 \hline
 \textit{less\_equal} & 0.17 & 30 & 0.17 \\
 \hline
 \textit{stdp\_case\_gen} & 0.34 & 66 & 0.60 \\
 \hline
 \textit{incdec} & 0.26 & 56 & 0.34 \\
 \hline
 \textit{stabilize\_func} & 0.12 & 158 & 0.36 \\
 \hline
 \textit{spike\_gen} & 1.46 & 28 & 1.55 \\
 \hline
 \textit{pulse2edge} & 0.44 & 22 & 0.44 \\
 \hline
 \textit{edge2pulse} & 0.49 & 58 & 0.61 \\
 \hline
 \end{tabular}}
 \label{tab:allmacros}
\end{table}

\section{Benchmarking and Results}
\label{benchmark}
This section presents 7nm post-synthesis power, performance, area (PPA) results for application-specific TNN prototypes.
% for the proposed custom macros 
% generated using Cadence toolchain. 
Performance is measured in terms of
% critical path delay for the individual macros, whereas for the TNN designs in Sections \ref{unsup_benchmark} and \ref{mnist_benchmark}, 
computation time (time taken to process one input), and is derived from the critical path delay and the gamma period as in \cite{nair2021online}. Area is the total cell and net area, while power includes dynamic (calculated using Cadence \textit{Joules}) as well as leakage power.

In order to demonstrate the efficacy of the TNN7 macros, we perform benchmarking for two groups of TNN prototype designs targeting
% based on the total synapse count using ASAP7 standard cells versus custom macros, for the benchmark column sizes reported in \cite{nair2020direct}, and then
two application domains: 
% 1) compare the layouts of two custom macros, namely, less\_equal and mux2to1gdi with corresponding ASAP7 standard cell-based modules to demonstrate the efficiency of our approach,
1) 36 single-column TNN designs for unsupervised time-series clustering on 36 UCR datasets
% , namely, \textit{largest} and \textit{smallest} with 6750 and 130 synapses respectively 
from \cite{chaudhary2021unsupervised}, with total synapse counts ranging from 130 to 6750;
% using the proposed custom macros as well as ASAP7-based standard cells and compare their PPA metrics with the corresponding 45nm values in Table IV in \cite{nair2020direct}, 
and 2) three much larger multi-layer TNN designs for MNIST digit recognition, namely, \textit{2-layer}, \textit{3-layer} and \textit{4-layer} TNNs (from \cite{smith2020temporal}) with total synapse counts of 389K, 1,310K and 3,096K, respectively.
% In addition to the methodology described in Section \ref{methodology}, 
% switching activity of 20\% 
% \textit{Joules} is incorporated into the \textit{Genus} synthesis flow for dynamic power calculations.
Following \cite{nair2021online}, an operating frequency of $100$ kHz is chosen for \textit{aclk} based on real-time operation requirement. We observe linear scaling of dynamic power with frequency and omit those results here for brevity.

\subsection{UCR Time-Series Clustering}
\label{unsup_benchmark}
As shown in \cite{chaudhary2021unsupervised}, TNN designs outperform or are competitive to state-of-the-art algorithms for unsupervised time-series clustering, averaging across the 36 UCR benchmark datasets. A specific column configuration is used for each of the 36 UCR datasets depending on the corresponding input size and number of clusters. While the hardware complexity analysis in \cite{chaudhary2021unsupervised} uses standard technology scaling to estimate the 7nm results from 45nm post-synthesis results, we present direct post-synthesis 7nm PPA results for all 36 TNN designs and further optimize them with our custom macros.

\begin{figure}[t]
\centering
\includegraphics[width=0.82\columnwidth]{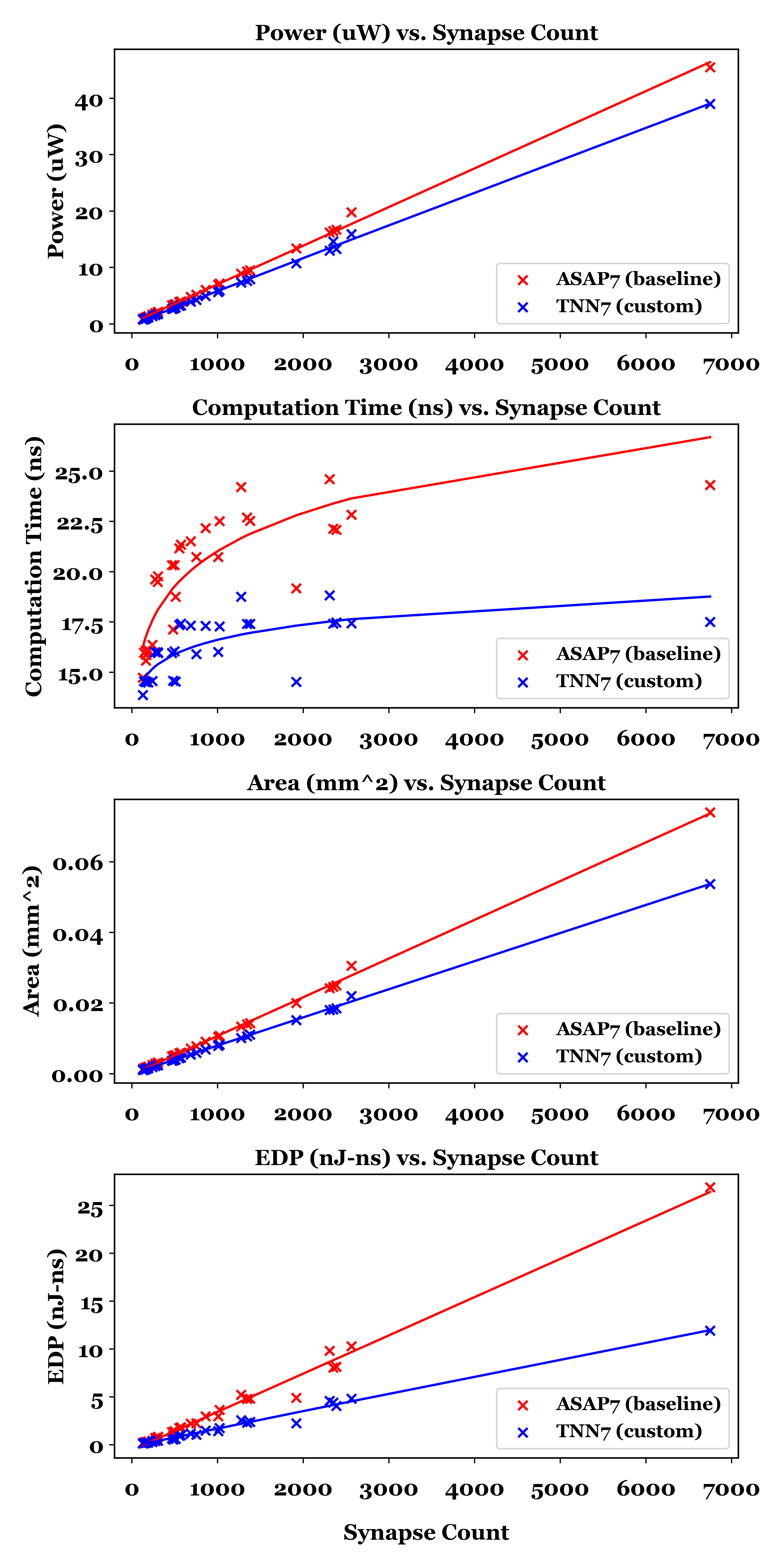} 
\caption{ASAP7 vs. TNN7 7nm PPA scaling across synapse counts for the 36 single column TNN designs as used in \cite{chaudhary2021unsupervised}}
\label{fig:date_ucr}
\end{figure}

To assess the range of PPA complexities for time-series clustering TNNs, we plot area, power, computation time, and energy-delay product (EDP), for the 36 single-column designs 
% used for unsupervised clustering on UCR time-series datasets, 
in Fig. \ref{fig:date_ucr}. EDP is used here to gauge both energy-efficiency and performance. 
%(JPS) The 36 \textit{p}x\textit{q} column sizes (out of which 6 are redundant) are listed to the right in Fig. \ref{fig:date_ucr}, for reference.
% Two important trends can be observed from this figure: 1) Area and power scale almost linearly with synapse count for both custom and standard designs, whereas computation time scales logarithmically, which corroborates the characteristic scaling equations in \cite{nair2021online}, and 2) The gap between standard and custom designs grows with increasing synapse counts, which implies, as TNN designs grow larger, they reap even more benefits from custom implementations.
Three key results can be observed here:
% from Fig. \ref{fig:date_ucr}:
\begin{enumerate}
    \item \textit{PPA synaptic scaling}: Area and power scale linearly with total synapse counts % (JPS) (\textit{p}*\textit{q}) 
    for both ASAP7 baseline and TNN7 custom designs, whereas computation time scales logarithmically with synapses per neuron (\textit{p}).
    This corroborates with the characteristic scaling equations in \cite{nair2021online}. 
    Note that x-axis is monotonic in \textit{p}*\textit{q} (not \textit{p}), making computation time data points non-monotonic in Fig. \ref{fig:date_ucr}.
    % making computation time non-monotonic. 
    % in Fig. \ref{fig:date_ucr}.
    % This scaling can be extended to any arbitrary TNN design.
    \item \textit{PPA improvements with TNN7}: TNN7 designs consume about 18\% less power and 25\% less area compared to baseline designs, and are about 18\% faster. EDP 
    % scales super-linearly with synapse count and 
    improves by more than 45\%,
    % due to custom macros. This 
    which clearly shows TNN7 designs are significantly more energy-efficient and are also faster. 
    % Note that even though two of the proposed macros have increased delays, the overall critical path in a column is still improved. 
    The gap between the two designs grows with increasing synapse count, which implies, as TNN designs grow larger, they reap even more benefits from custom macros.
    \item \textit{Potential for low-power edge-native sensory processors}: With custom macros, even the \textit{largest} TNN column with 6,750 synapses consumes just 0.054 mm\textsuperscript{2} area and 39 $\mu$W power. Note that this also accounts for on-chip learning via STDP, highlighting the value of proposed macros for highly energy-efficient TNN sensory processing units capable of online continuous learning.
\end{enumerate}

\subsection{MNIST Digit Recognition}
\label{mnist_benchmark}
Here, we move to much larger TNN designs and evaluate three multi-layer TNN prototypes for MNIST digit recognition, with different design points in the error rate vs. hardware complexity tradeoffs. The three designs are as follows:
% illustrated in Fig. \ref{fig:brookes}. The three prototype designs are: 
1) 2-layer TNN (389K synapses and 7\% error) derived from ECVT in \cite{smith2020temporal}; 2) 3-layer TNN (1.31M synapses and 3\% error) derived from ECCVT in \cite{smith2020temporal}; and 3) 4-layer TNN (3.096M synapses and 1\% error) derived from ECCCVT in \cite{smith2020temporal}. Table \ref{tab:tnn_img} provides 7nm PPA for these designs, derived using synaptic count scaling as in \cite{nair2021online}.
% Note that TNN7 is specific to the ``C'' layers and does not yet support the ``\textit{VT}'' layers \cite{smith2020temporal} that are a simpler form of TNN columns.
Note that ``C'' layers above consist of TNN7 columns, however the ``VT'' layers \cite{smith2020temporal}, that are a simpler form of TNN columns, are currently not supported within TNN7.
Hence, the synaptic scaling here treats all network layers as ``C'', thereby providing an upper limit on the PPA complexity.
% by assuming all network layers to be composed of TNN columns as in Fig. \ref{macro_bd}.
% , whereas as per \cite{smith2020temporal}, the last layer consists of components that are microarchitecturally different from TNN columns, with lesser hardware complexity.
% Note that the power values are in mW and area in mm\textsuperscript{2}.
% 
% \begin{figure}[t]
% \centering
% \includegraphics[width=\columnwidth]{mnist_linear_mod.png} 
% \caption{Power vs. MNIST error for standard vs. custom implementations of three multi-layer TNN prototypes in \cite{smith2020temporal}}
% \label{fig:brookes}
% \end{figure}
% 
\begin{table}[t]
\centering
\caption{ASAP7 vs. TNN7 7nm PPA comparison for three TNN prototype designs for MNIST from \cite{smith2020temporal}}
% \scalebox{0.97}{
\resizebox{\columnwidth}{1.43cm}{
 \begin{tabular}{|c|c|c|c|c|c|c|} 
 \hline
 TNN & Synapse & Error & Cell & Power & Comp. & Area \\
 Design & Count & Rate & Library & ($mW$) & Time ($ns$) & ($mm^{2}$)\\
 \hline
 \hline
 \multirow{2}{*}{2-Layer} & \multirow{2}{*} {389K} & \multirow{2}{*} {7\%} & ASAP7 & 2.62 & 49.00 & 4.27 \\ 
 \cline{4-7}
 & & & TNN7 & 2.25 & 41.38 & 3.09 \\
 \hline
 \hline
 \multirow{2}{*} {3-Layer} & \multirow{2}{*} {1,310K} & \multirow{2}{*} {3\%} & ASAP7 & 8.83 & 78.37 & 14.37 \\ 
 \cline{4-7}
 & & & TNN7 & 7.57 & 66.16 & 10.42 \\
 \hline
 \hline
 \multirow{2}{*} {4-Layer} & \multirow{2}{*} {3,096K} & \multirow{2}{*} {1\%} & ASAP7 & 20.86 & 108.46 & 33.95 \\ 
 \cline{4-7}
 & & & TNN7 & 17.89 & 91.58 & 24.63 \\
 \hline
 \end{tabular}
 }
  \label{tab:tnn_img}
\end{table}

%% hey! how does adding the trendline look? Hmmm, the trendline should probably follow the log curve. Try to fit a polynomial instead of linear Are you using MATLAB?
% No python! I see. If you give me the numbers, I can do a polynomial trendline fit for you in Matlab. I can also share my Matlab code with you if you like?
% how about i give you access to my notebook file? Sure. It currently looks good but there seems to be a sharp corner - can we make it smoother with higher order polynomial?
% 
% 
% \begin{table}[t]
% \centering
% \caption{Standard vs. custom 7nm PPA comparison for three TNN prototype designs for MNIST from \cite{smith2020temporal}}
% \scalebox{0.95}{
%  \begin{tabular}{|c|c|c|c|c|c|c|} 
%  \hline
%  TNN & Synapse & Error & Cell & Power & Comp. & Area \\
%  Design & Count & Rate & Library & ($mW$) & Time ($ns$) & ($mm^{2}$)\\
%  \hline
%  \hline
%  \multirow{2}{*}{2-Layer} & \multirow{2}{*} {389K} & \multirow{2}{*} {7\%} & Standard & 2.62 & 45.94 & 4.27 \\ 
%  \cline{4-7}
%  & & & Custom & 2.25 & 38.79 & 3.09 \\
%  \hline
%  \hline
%  \multirow{2}{*} {3-Layer} & \multirow{2}{*} {1,310K} & \multirow{2}{*} {3\%} & Standard & 8.83 & 51.11 & 14.37 \\ 
%  \cline{4-7}
%  & & & Custom & 7.57 & 43.15 & 10.42 \\
%  \hline
%  \hline
%  \multirow{2}{*} {4-Layer} & \multirow{2}{*} {3,096K} & \multirow{2}{*} {1\%} & Standard & 20.86 & 54.23 & 33.95 \\ 
%  \cline{4-7}
%  & & & Custom & 17.89 & 45.79 & 24.63 \\
%  \hline
%  \end{tabular}\\
%  }
% %  \vspace{2 mm}
%   \label{tab:tnn_img}
% \end{table}
% 
From Table \ref{tab:tnn_img}, similar PPA improvements with custom macros 
%as in Section \ref{unsup_benchmark} 
can be observed for these complex multi-layer TNNs (14\%, 16\%, and 28\% improvements on power, performance and area, respectively). 
% A 4-layer TNN with 3M synaptic weights can achieve upto 99\% accuracy on MNIST and is capable of online continuous learning \cite{smith2020temporal}, \cite{nair2021online}. 
The 4-layer TNN with 3M synaptic weights and 99\% MNIST accuracy consumes only 17.89 mW power and 24.63 mm\textsuperscript{2} area. 
% can have a peak throughput of about 11M images per second.
% process a new MNIST input every 91.58 ns. 
% We use MNIST only as an illustration tool due to its ubiquity. 
% The 4-layer TNN design here 
This TNN represents an edge-native real-time sensory processing unit that is capable of both online (MNIST-like) image-based classification and continuous learning, while consuming less than 20 mW power. 
% Fig. \ref{fig:brookes} graphically depicts the power consumption vs. MNIST error rate tradeoff observed in TNNs, along with Minerva \cite{reagen2016minerva} for reference. 
% It also plots Minerva from Fig. 1 in \cite{reagen2016minerva} for reference. 

Using the survey of MNIST neural networks from \cite{reagen2016minerva}, it can be observed that for similar accuracies, TNN-based processing units that consume a few tens of mW power are about 1000x more efficient comparing to GPUs, 100x comparing to FPGAs and 10x comparing to many state-of-the-art 
ASICs that consume a few hundreds of mW power. 
Furthermore, TNN7 enhances this scalability as it offers a lower-cost trajectory in the accuracy vs. hardware complexity tradeoff.
\section{Synthesis Runtime Evaluation}
A further advantage to using a custom cell library is 
% improved synthesis runtime for 
significantly faster design netlist generation.
% compared to utilizing the standard cell library. 
As the macro design instances are preserved and not manipulated during synthesis, it enables the synthesis tool to realize a design hierarchy by directly mapping the hard macros, thereby mitigating the combinatorial search space complexity for the optimization tool. 
% A considerable portion of the synthesis speedup is realized during the mapping and optimization phases of the TNN7-based column synthesis.
To evaluate this benefit for TNN7, we use the following setup:
% The setup for this experiment is described as follows. 
Genus v19.1 is run on a server comprising of 48 Intel(R) Xeon(R) E5-2680 CPU cores with
% with a maximum operating frequency of 2.5GHz. 
the maximum number of CPUs utilized set to 8. Synthesis was performed on the same column configurations from Section \ref{unsup_benchmark}, with the TNN7 custom macros as well as without them (ASAP7 baseline). 
% A bash script was used to automate the synthesis runs in the manner of standard cell-based synthesis immediately followed by a custom macro-based synthesis run for a given configuration.

\begin{figure}[t]
    \centering
    \includegraphics[width=\columnwidth]{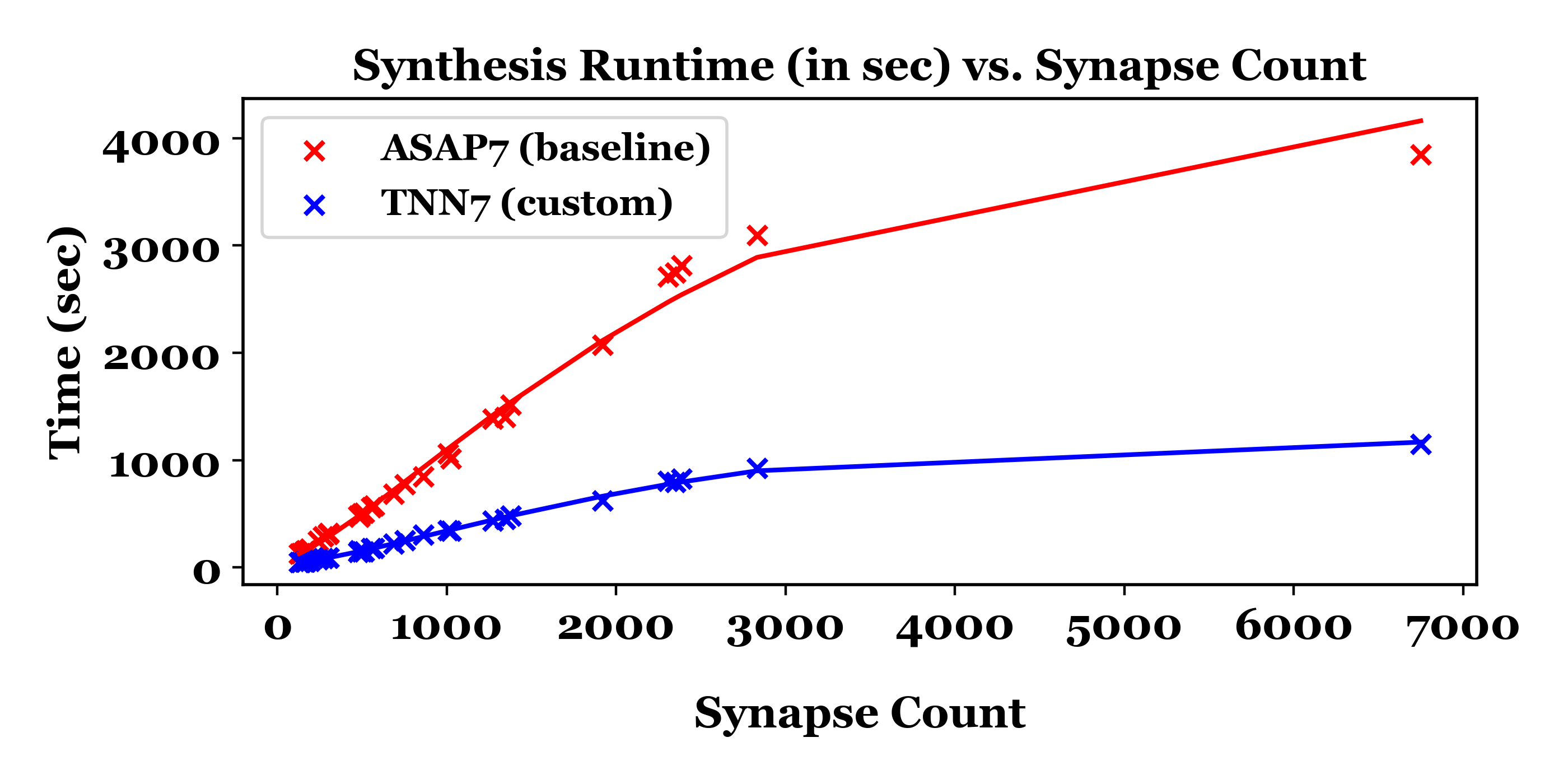}
    \caption{ASAP7 vs. TNN7 synthesis runtime comparison}
    \label{synthrun}
\end{figure}

Fig. \ref{synthrun} depicts the runtimes for both standard ASAP7-based and corresponding TNN7-based designs.
%%HELLOOOO - Heyyyy
%% how is the paper looking? Can i do something in specific - Yes, Fig 10 and 11 - can you make the fonts bigger and bolder? And align Fig 11 as Power, Perf, Time and EDP
%% got it, I can do that. I will so see if I can write synthesis runtime better - I think it looks good - currently doing some edits - feel free to edit it after I am done - but more immportantly section III - can we make it better? Hmm, I can look into that , let me think about it - Great thanks! I am finished with Sections I/II, IV and now working on V. Also added Fig 1 (modified) - let me know what you think
%% Yeah i saw the figure, looks great! And awesome, thanks for editing those sections :D Is there a better polygon to use instead of the tilted rhombus for synapse in Fig 1? I couldnt think of a better shape :P Hmmm very good question haha, I will hgave to refer to draw.io and see, can look into that - yeah maybe at the end later sure, will do! Thanks!
On average, TNN7 speeds up the netlist generation (including mapping and optimization) by 3.17x with respect to the baseline ASAP7-based designs. Using TNN7, the largest column with 6750 synapses is synthesized in 926 seconds ($\sim$15 minutes), as opposed to 3849 seconds ($\sim$1 hour) for the baseline design. Fig. \ref{synthrun} illustrates increasing runtime benefits for TNN7 as the designs grow larger. This trend can be extrapolated beyond single columns to multi-layered networks based on synapse counts, demonstrating the scalability of TNN7 to realize deep TNNs, that would have otherwise suffered from long runtimes.
% Averaged across 32 different synapse counts, there is a 3.17x improvement in synthesis runtime when generating the netlist model for TNN7-based column design compared to the ASAP7-based column design. For the largest column configuration of 270x25, runtime for TNN7-based design was completed in 00:15:26 (hh:mm:ss format) as opposed to 01:04:09 for ASAP7-based design.  
% This improved runtime allows for netlist generations of highly-scaled column configurations or deep-layered TNN frameworks, which would have otherwise suffered from long runtimes. 
% 
% 
% \vspace{-10pt}
\section{Concluding Remarks}
Prior works 
% \cite{smith2020temporal, nair2021online, chaudhary2021unsupervised} 
have shown that TNNs can achieve highly energy efficient brain-like sensory processing. This work develops a customized 7nm cell library, TNN7, consisting of nine new macros to enable extensive TNN design optimization. The TNN7 macros yield 14\%, 16\%, 28\% and 45\% improvements in power, performance, area and EDP, respectively. This surpasses typical area-power-delay trade-offs by achieving significant improvements in all three PPA metrics. With TNN7, competitive performance to state-of-the-art can be achieved on time-series clustering with just 40 $\mu$W and 0.05 mm\textsuperscript{2}, and on MNIST with
% With the larger 4-layer TNN design (3M synaptic weights), we can achieve upto 99\% accuracy on MNIST while consuming 
only 17.89 mW and 24.63 mm\textsuperscript{2}. This shows the feasibility of TNN-based edge-native 
neuromorphic processors capable of online continuous learning.
% 
% a competitive single-column TNN prototype for time-series only consumes 21.6 $\mu$W power and 0.464 mm\textsuperscript{2} area, and a 4-layer MNIST TNN (with 99\% accuracy) consumes 
% a 2-layer TNN prototype with 13,750 neurons and 315,000 synapses can be implemented in 7nm CMOS. 
% 
% Using the custom macros, significant reduction in all three PPA metrics can be achieved. This transcends the typical tradeoff that exists between them, where improving one metric can potentially exacerbate the other metrics. 
% Furthermore, it enables scalable TNN designs with potential for multiple orders of magnitude improvements in power consumption as compared to conventional DNN hardware.
% 
% We plan to extend this work to a test chip tape-out in the near future.
\begin{figure}[t]
    \centering
    \begin{subfigure}[b]{0.47\columnwidth}
        \centering
        \includegraphics[width=\columnwidth]{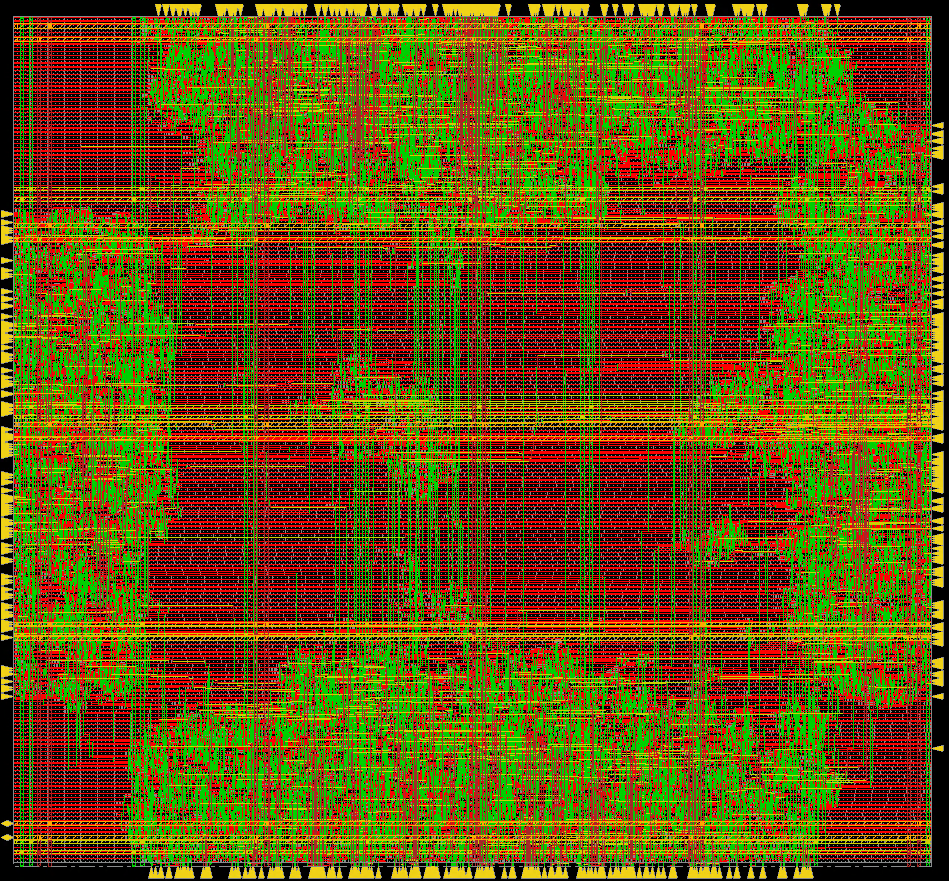}
        \caption{ASAP7}
        \label{stdlayout}
    \end{subfigure}%
    \hspace{1pt}
    \begin{subfigure}[b]{0.23\textwidth}
        \centering
        \includegraphics[width=\columnwidth]{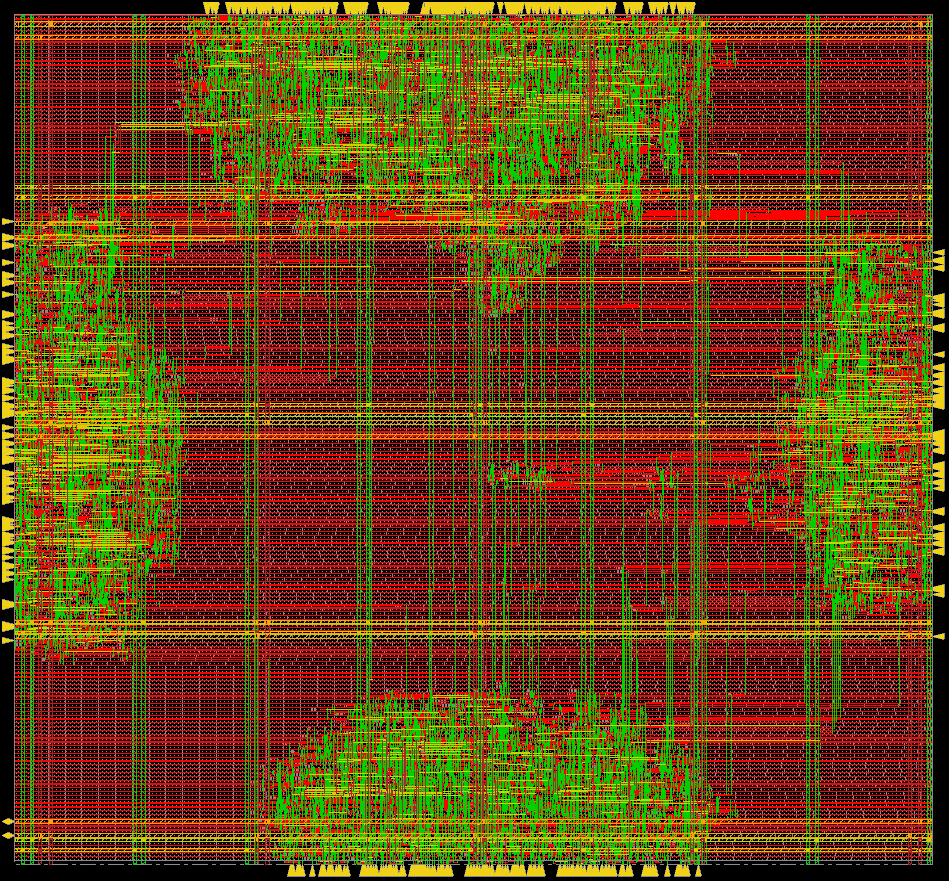}
        \caption{TNN7}
        \label{cuslayout}
    \end{subfigure}
    \label{clayout}
    \caption{ASAP7 vs. TNN7 layouts for \textit{82}x\textit{2} column
    % developed for \textit{TwoLeadECG} application from \cite{chaudhary2021unsupervised}
    % \cite{UCRArchive2018}
    }
\end{figure}

This work can serve as a foundation for building a complete design framework and toolsuite, 
% for application-specific prototyping of TNNs
that can translate application-specific TNN designs from the functional level (software models) to hardware implementation and physical design. Towards that goal, our ongoing work involves open-sourcing\footnote{https://github.com/prabsy96/TNN7} the custom macros and developing an automated RTL-to-GDSII process flow, to generate signoff layout and PPA metrics for arbitrary TNN designs. Fig. 13 illustrates both baseline and TNN7-based place-and-route layouts for 
% one of the 36 TNN designs from Section \ref{unsup_benchmark}, specifically, 
the \textit{82}x\textit{2} column developed for UCR \textit{TwoLeadECG} application
% from the UCR archive
as used in \cite{chaudhary2021unsupervised}.
% eventually targeting a chip tapeout. 
The layouts corroborate the efficacy of the proposed macros as the routing density in the custom design (Fig. \ref{cuslayout}) is visibly less complex as compared to the baseline design (Fig. \ref{stdlayout}).
% We believe brain-like edge-native sensory processing units can be effectively implemented in standard digital CMOS that consume on the order of 10 mW of power and 100 mm\textsuperscript{2} of die area. 
Furthermore, the custom library can be generalized to include the space-time primitives in \cite{smith2018space} and thereby implement any bounded space-time function directly in CMOS.

% This paper lays the foundation for our ongoing work on building an automated TNN design framework for pushing application-specific TNN configurations through the RTL-GDSII process flow in an automated manner, giving end-users using this toolchain ability to realized PPA-specific optimized process runs. For example, the framework will enable the users to push any of the 36 TNN standard or custom based designs throught the process flow to obtain signoff layout (Figure 12 depecits the TwoLeadECG TNN model place & route layout in ASAP7) and PPA values, with the help of a few command line options. We will further incorporate MNIST -specific multi-layer TNN models that can leverage both standard and custom-based columns to perform MNIST unsupervised learning.  

\bibliographystyle{IEEEtranS}
\bibliography{refs}

\end{document}